\documentclass{revtex4}

\usepackage[dvips]{graphicx}

\usepackage{amssymb,amsfonts,amsmath,epsfig}

\begin{document}

%%%%%%%%%%%%%%%%%%%%%%%%%%%%%%

%% For titles, only capitalize the first letter
%% \title{Almost sharp fronts for the surface quasi-geostrophic equation}

\title{Serial correlation and heterogeneous volatility in financial markets: beyond the LeBaron effect}

%% Enter authors via the \author command.  
%% Use \affil to define affiliations.
%% (Leave no spaces between author name and \affil command)

%% Note that the \thanks{} command has been disabled in favor of
%% a generic, reserved space for PNAS publication footnotes.

%% \author{<author name>
%% \affil{<number>}{<Institution>}} One number for each institution.
%% The same number should be used for authors that
%% are affiliated with the same institution, after the first time
%% only the number is needed, ie, \affil{number}{text}, \affil{number}{}
%% Then, before last author ...
%% \and
%% \author{<author name>
%% \affil{<number>}{}}

%% For example, assuming Garcia and Sonnery are both affiliated with
%% Universidad de Murcia:
%% \author{Roberta Graff\affil{1}{University of Cambridge, Cambridge,
%% United Kingdom},
%% Javier de Ruiz Garcia\affil{2}{Universidad de Murcia, Bioquimica y Biologia
%% Molecular, Murcia, Spain}, \and Franklin Sonnery\affil{2}{}}

\author{Simone Bianco}  
  
\affiliation{Department of Applied Science, College of
    William and Mary, Williamsburg, Va 23187-8795, USA}
\author{Fulvio Corsi}
\author{Roberto Ren\`o}
\affiliation{Dipartimento di Economia Politica, Universit\`a di
    Siena, Piazza S.Francesco 7, 53100, Siena, Italy}
%% The \maketitle command is necessary to build the title page.

%%%%%%%%%%%%%%%%%%%%%%%%%%%%%%%%%%%%%%%%%%%%%%%%%%%%%%%%%%%%%%%%

\begin{abstract}
We study the relation between serial correlation of financial returns and
volatility at intraday level for the S$\&$P500 stock index. 
At daily and weekly level, serial correlation and volatility are
known to be negatively correlated (LeBaron effect). While confirming that the
LeBaron effect holds also at intraday 
level, we go beyond it and, complementing the efficient market
hyphotesis (for returns) with the heterogenous market hyphotesis (for
volatility), we test the impact of unexpected volatility, defined
as the part of volatility which cannot be forecasted, on the presence of
serial correlations in the time series. 
We show that unexpected volatility is instead positively correlated
with intraday serial correlation.
\end{abstract}
\maketitle

%% When adding keywords, separate each term with a straight line: |

%% Optional for entering abbreviations, separate the abbreviation from
%% its definition with a comma, separate each pair with a semicolon:
%% for example:
%% \abbreviations{SAM, self-assembled monolayer; OTS,
%% octadecyltrichlorosilane}

%% The first letter of the article should be drop cap: \dropcap{}
%\dropcap{I}n this article we study the evolution of ''almost-sharp'' fronts

%% Enter the text of your article beginning here and ending before
%% \begin{acknowledgements}
%% Section head commands for your reference:
%% \section{}
%% \subsection{}
%% \subsubsection{}

Serial correlation of asset prices
is one of the most elusive quantity of financial
economics. According to the theory of efficient markets \cite{Fam70,DoyLo99},
it should not exist at all, and when it exists it represents an
anomaly of financial markets.
Many economists and physicist
devoted themselves to the study of stock return predictability
\cite{DoyPatZov05,PinKal04}. Historical returns should prevent any
forecasting technique, even if it has been shown, like in
\cite{LoMac88}, that the
random walk hypothesis holds only weakly.

On the other hand the variance of financial returns
on a fixed time interval, which is called volatility, is
a highly predictable quantity~\cite{Eng82, Bol86}, 
with its probability distribution
function showing fat tails~\cite{Sta97}. The natural association of volatility to
financial risk forecast and control makes its analysis paramount in
Economics. To some extent, it seems obvious therefore to link volatility to returns serial
correlation. If anything else, the link between volatility and serial 
correlations can reveal basic properties of the price formation mechanism. 

A notable stylized fact on serial correlation is
the LeBaron effect \cite{LeB92}, according to which  
volatility is negatively correlated to serial correlation, relation that  has
been empirically confirmed at daily level on several
markets~\cite{VenPee05}. We exploit, as in \cite{BiaRen06,BiaRen08},
high-frequency information (that is, intraday data) 
to test this effect more precisely than in previous studies. 
With respect to \cite{BiaRen06,BiaRen08}, not only we
extend considerably our data set, but also 
our work is set within the framework of the Heterogenous Market
Hypothesis~\cite{Muletal93}, which implies the modification of some basic
assumptions on financial market dynamics due to empirical
findings. The main feature of this model is the assumption that agents must
be considered heterogenous, namely, they react differently to incoming news.  
This implies a completely new interpretation of the concept of time in the
market, as now every agent has its own dealing time and frequency. More
importantly, the transactions time, in the agent framework, has to be related
to short-, medium-, and long-term decisions. Thus, volatility will be itself
consistently composed by a cascade of several time components. 
In this work,
the heterogeneous market hypothesis is the basis of 
our method of volatility estimation, introduced for the first time
in~\cite{Cor04}. A notable advantage in modeling volatility 
in the context of heterougeneous markets is that it allows to reconcile the measure 
with some empirical findings on volatility (e.g., long-range
dependence and fat tails), a feature which simple autoregressive 
volatility forecasting models lacks. 

The purpose of this paper is then to exploit intraday 
measures of volatility and serial correlation to investigate the
LeBaron effect at a deeper level with a very large and liquid dataset.
We find that the LeBaron effect holds. However, we also refine the
finding of LeBaron by showing that volatility can be splitted into two
components: a predictable one and an unpredictable one. While the
first is negatively correlated with serial correlation, the latter is
instead positively correlated with serial correlation.

\section{Methods and data description}
Let us assume to have 
$r_1,\ldots ,r_n$ intraday logarithmic returns. 
To quantify volatility, we construct daily {\it realized volatility}
measures defined as the cumulative sum of squared intraday
five-minutes returns \cite{AndBolDie03}:
\begin{equation}
RV_t = \sum_{i=1}^{n} r^2_{i,t}
\end{equation} 
where $r_{i,t}$ is the $i$-th return in day $t$. Since volatility has been
proved to be log-Normal~\cite{CizEtAl97}, we use the logarithm of $RV_t$ to obtain Normal
distributions.

Measuring daily serial correlation is more subtle, and we borrow from
\cite{BiaRen08} using a modified overlapped Variance Ratio.
Define:
\begin{eqnarray}
  \hat{\mu} \equiv \frac{1}{n}\sum_{k=1}^n r_k \\
  \hat{\sigma}_a^2 \equiv \frac{1}{n-1} \sum_{k=1}^{n} (r_k - \hat{\mu})^2\\
  \hat{\sigma}_c^2 \equiv \frac{1}{m}  \sum_{k=q}^{n}
  \left(\sum_{j=k-q+1}^k r_j - q\hat{\mu}\right)^2
\end{eqnarray}
where 
\begin{equation}
  m = q(n - q + 1)\Bigg(1 - \frac{q}{n}\Bigg).
\end{equation}
We define the variance ratio as follows:
\begin{equation}\label{vrover}
 VR(q) = \left( \frac{\hat{\sigma}_c^2}{\hat{\sigma}_a^2} \right)^\beta.
\end{equation}
The use of the power transformation  $f(x)=x^\beta$ makes the
distribution closer to a Normal one in small samples \cite{CheDeo06}. 
The expression of Eq.~\eqref{vrover} is asymptotically Normal with mean $1$
and defined standard deviation. 
$\beta$ is given by:
\begin{equation}
\beta =
1-\frac23\frac{\left(\sum_{j=1}^{(n-1)/2}W_k(\lambda_j)\right)\left(\sum_{j=1}^{(n-1)/2}W^3_k(\lambda_j)\right)}{\left(\sum_{j=1}^{(n-1)/2}W^2_k(\lambda_j)\right)^2},
\end{equation}
where $W_k$ is the Dirichlet kernel:
\begin{equation}
W_k(\lambda) = \frac 1k \frac{\sin^2(k\lambda/2)}{\sin^2(\lambda/2)}
\end{equation}
and $\lambda_j=2\pi j/n$.

Intuitively, the variance ratio expresses the ratio of variances
computed at two different frequencies, whose ratio is given by $q$. If
there is no serial correlation in the data, $VR(q)$ should be close to
one. In presence of positive serial correlation, the variance computed at
at the higher frequency is lower, and $VR(q)<1$. If instead there is
negative serial correlation, this argument reverts and $VR(q)>1$. 
We
use $q=2,3,4,5,6$. Higher values of $q$ cannot be used without
introducing possible distortions in the statistics behavior \cite{DeoRic03}. 
The $VR$ measure has been shown to be correct also for
heteroskedastic data generating process~\cite{LoMac88}, and it is
defined with overlapping
observations~\cite{DeoRic03}. This measure is a reliable measure of serial
correlation both at daily~\cite{LoMac88} and intraday~\cite{BiaRen06,BiaRen08}
level.

The data set we use is one of the most liquid financial asset in the world,
that is the S\&P500 stock index futures from 1993 to 2007, for a total
of $4344$ days. We have all high-frequency information, but to avoid for
microstructure effects we use a grid of $84$ daily 5-minutes logarithmic
returns,
interpolated according to the previous-tick scheme (the price at time $t$
is the last observed price before $t$). These choices are the standard
in this kind of applications.

Relation between volatility and serial correlation can simply be tested
using a linear regression 
\begin{equation}\label{simple}
VR(q)_t = b + c\log RV_{t} + \varepsilon_t .
\end{equation}

However, the LeBaron effect tests the relation between serial
correlation and {\it lagged} volatility, which can be tested via the
regression:
\begin{equation}\label{lagged}
VR(q)_t = b + c_0\log RV_{t} + c_1\log RV_{t-1}  +  \varepsilon_t .
\end{equation}

If we want to add further lagged values of realized volatility, we
cannot ignore the fact that volatility is well known to display
long-range dependence. One effective way to accommodate for this
stylized fact without resorting to the estimation burden of a long
memory model is the HAR model of \cite{Cor04}.
%instead to introduce heterogeneity in the
%volatility components \cite{Muletal90}. This can be done with the HAR model of \cite{Cor04}.
Following what can be termed the "Heterogeneous Market Hypothesis" of
\cite{Dacetal01, LuxMar99, LeB02, WerWer00,McMSpe06} 
which recognizes the presence of heterogeneity in traders horizon and 
the asymmetric propagation of volatility cascade from long to short time periods \cite{Ghaetal96},
the basic idea that emerges is that heterogeneous market structure generates an heterogeneous volatility cascade.
Hence, \cite{Cor04} proposed a stochastic additive cascade of three different realized volatility components 
which explain the long memory observed in the volatility as the superimposition
of few processes operating on different time scales corresponding 
to the three typical time horizons operating in the financial market: daily, weekly and
monthly. 
This stochastic volatility cascade leads to a simple AR-type model in the 
realized volatility with the feature of considering realized volatilities defined over heterogeneous
time periods (the HAR model):
\begin{eqnarray}\label{HAR}
\log RV_t &=& \beta_0+ \beta_{(d)}\log RV_{t-1} \nonumber \\
 & + & \beta_{(w)}\log RV^{(w)}_{t-1} + \beta_{(m)}\log RV^{(m)}_{t-1} + \eta_t
\end{eqnarray}
where $\eta_t$ is a zero-mean estimation error and:
\begin{equation}
\log RV^{(w)}_{t-1} = \frac 15 \sum_{k=1}^5 \log RV_{t-k},~~~~~~~\log RV^{(m)}_{t-1} = \frac 1{22} \sum_{k=1}^{22} \log RV_{t-k}.\end{equation}
Although the HAR model doesn't formally belong to
the class of long-memory models, it generates apparent power laws and
long-memory, i.e. it is able to reproduce a memory decay which is 
indistinguishable from that observed in the empirical data. It is now
widely used in applications in financial economics~\cite{AndBolDie07,GhySanVal06}. 
We estimate the HAR model with ordinary least squares, and use the estimated
coefficients $\hat{\beta}_{0,(d),(w),(m)}$
define the predictable volatility as:
\begin{equation}\sigma_{p,t} = \hat{\beta}_0+ \hat{\beta}_{(d)}\log RV_{t-1} +
\hat{\beta}_{(w)}\log RV^{(w)}_{t-1} + \hat{\beta}_{(m)}\log
RV^{(m)}_{t-1}\end{equation}
and {\it unexpected volatility} as the residuals of the above
regression:
\begin{equation}\sigma_{u,t} = \hat{\eta}_t,\end{equation}
allowing us to estimate the linear model:
\begin{equation}\label{up}
VR(q)_t = b + c_u\sigma_{p,t} + c_p\sigma_{u,t}+ \varepsilon_t 
\end{equation}
which fully takes into account heterogeneity, long-range dependence and
heteroskedasticity of financial market volatility.

\section{Results and discussion}
\begin{figure*}
\mbox{\epsfig{figure=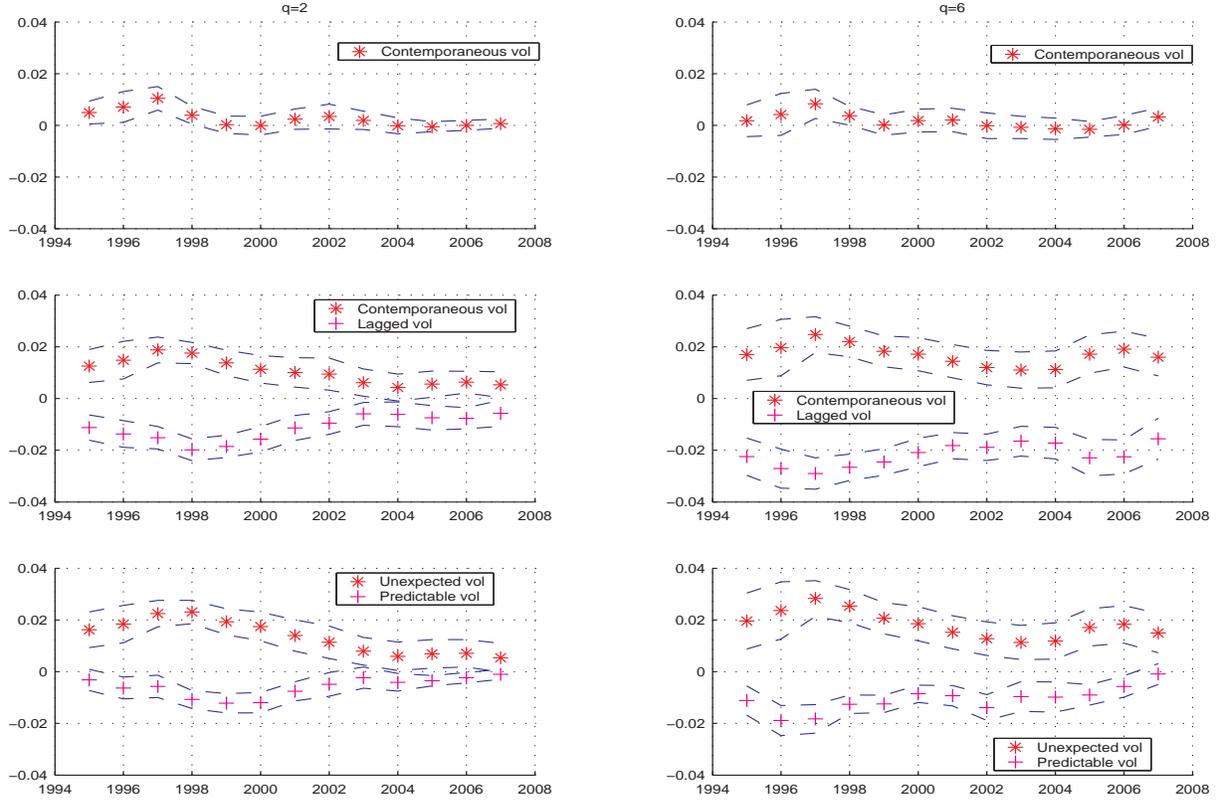,width =0.9\linewidth,height=0.6\linewidth}}
\caption{On a rolling window of lenght $5$ years, shows: (top) the
  measures of coefficient $c_0$ of regression (\ref{simple}); (middle)
the measures of coefficients $c_0$ and $c_1$ of regression
(\ref{lagged}); (bottom) the measures of coefficients $c_0$ and $c_1$
of regression (\ref{up}). On the left column, the Variance ratio have
been computed with $q=2$ (5 minutes); on the right column, with $q=6$
(25 minutes). Dashed lines represents confidence intervals at $95\%$
confidence level. This figure shows that the correlation between
serial correlation and predictable volatility is negative (LeBaron
effect), while the correlation between serial correlation and
unexpected volatility is positive.
}\label{mainfigure}
\end{figure*}

We estimate the regression [\ref{simple}] on a rolling window five
years long
(Figure \ref{mainfigure}, with $q=2,6$) as well as on the
%whole sample (Table \ref{maintable}, $q=2,3,4,5,6$).
whole sample (Table 1, $q=2,3,4,5,6$).
Results in the first row of Figure \ref{mainfigure} may look
disappointing, since the LeBaron effect is not displayed when looking
just at the correlation between realized volatility and variance
ratio. Moreover, this correlation tends to be slightly positive
instead of negative, especially in the first part of the sample.

However, LeBaron evidenced the negative correlation between $r^2_t$ and
$r_tr_{t+1}$.
This is indeed what the second row of
Figure \ref{mainfigure} shows: the coefficient of lagged volatility on
variance ratio is negative and significant across all the sample. 

\begin{figure*}
\mbox{\epsfig{figure=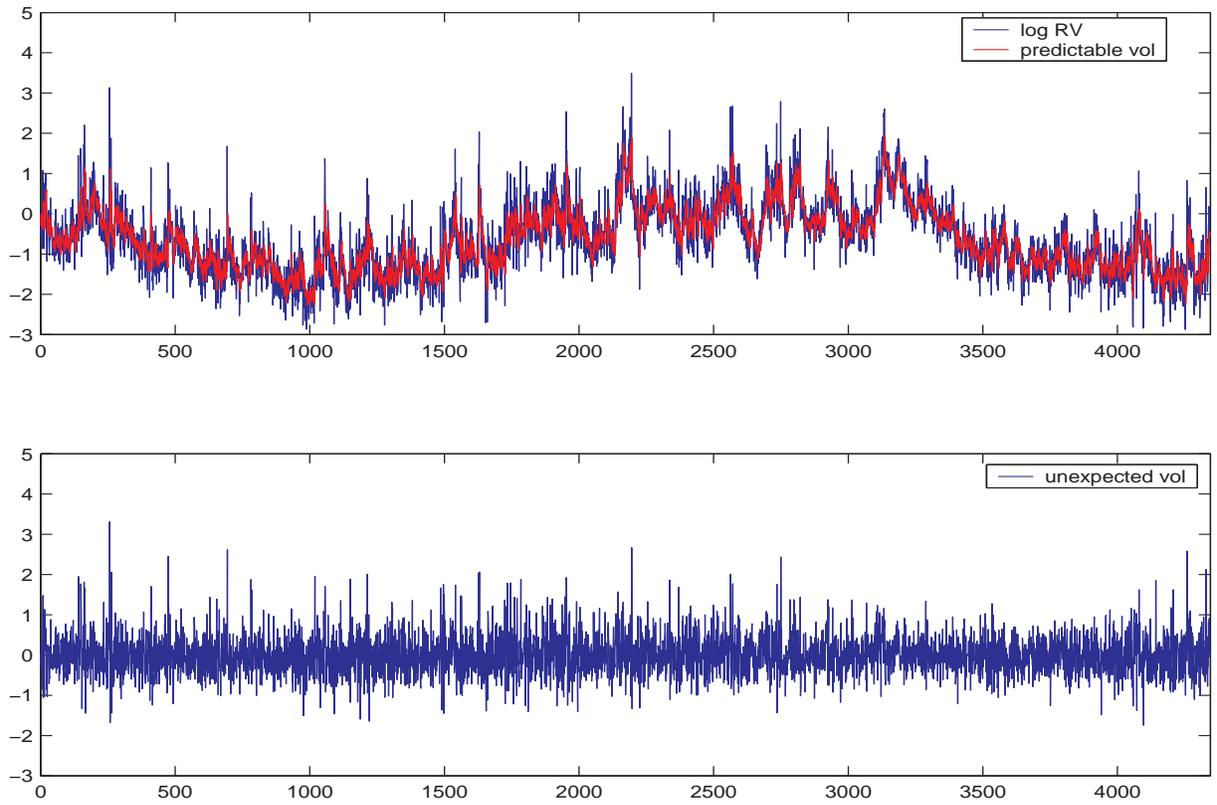,width
    =0.9\linewidth,height=0.6\linewidth}}
\caption{Top: shows the time series estimate of $\log RV_t$ used in
  this study ($4344$ observations), together with the predictable
  volatility estimated by means of the HAR model. Bottom: shows the
  time series of unexpected volatility estimated as the residuals
  of the HAR model.}\label{HARfigure}
\end{figure*}

Most interestingly, when the LeBaron effect is correctly disentangled in
the data, we find that contemporanoues volatility is significantly and
positively correlated with the variance ratio. 
Hence, estimation results for equation [\ref{lagged}] indicate a sharp difference 
in the relation between serial correlation and volatility: strongly
positive for contemporaneous volatility and strongly negative for
lagged one. 
Such antithetical behavior of the relation is even more puzzling
considering the well known stylized fact of volatility to be highly
persistent.  How could we explain this result? 
By our heterogeneous ``rotation'' 
of the regressors, we can rewrite Equation [\ref{lagged}] in the form
[\ref{up}]. This provides the separation between predictable and
unexpected volatility illustrated in Figure \ref{HARfigure}.
The new specification greatly helps in shedding light
on this result providing a precise economic interpretation. 
Hence, as \cite{BiaRen08} suggested, we can now provide an
explanation in term of predictable and unexpected volatility:
since volatility is known to be predictable by
market participants, it has a different impact with respect to its
unpredictable component.

The third row of Figure
\ref{mainfigure} shows indeed that the predictable volatility, defined
by means of the HAR model, is negatively correlated with the variance
ratio (more with higher $q$) and that the unpredictable volatility is
positively correlated with the variance ratio (more with higher
$q$). 

%The full sample estimates in Table \ref{maintable} corroborate this finding. 
The full sample estimates in Table 1 corroborate this finding. 
We can then rephrase the LeBaron effect as follows: serial
correlation is negatively correlated with the expected
volatility. Moreover, we can conclude that serial correlation is
instead positively correlated with unexpected volatility, which is a
previously unrecognized stylized fact of financial returns.
This stylized fact suggests that the usual explanation of the LeBaron
effect in terms of feedback trading \cite{SenWad92} is at least
incomplete, advocating for a broader theory  
on the link between volatility and
the way information is spread to heterogenous market components.

%% == end of paper:

%% Optional Materials and Methods Section
%% The Materials and Methods section header will be added automatically.

%% Enter any subheads and the Materials and Methods text below.
%\begin{materials}
% Materials text
%\end{materials}

%% Optional Appendix or Appendices
%% \appendix Appendix text...
%% or, for appendix with title, use square brackets:
%% \appendix[Appendix Title]

%% PNAS does not support submission of supporting .tex files such as BibTeX.
%% Instead all references must be included in the article .tex document. 
%% If you currently use BibTeX, your bibliography is formed because the 
%% command \verb+\bibliography{}+ brings the <filename>.bbl file into your
%% .tex document. To conform to PNAS requirements, copy the reference listings
%% from your .bbl file and add them to the article .tex file, using the
%% bibliography environment described above.  

%%  Contact pnas@nas.edu if you need assistance with your
%%  bibliography.

% Sample bibliography item in PNAS format:
%% \bibitem{in-text reference} comma-separated author names up to 5,
%% for more than 5 authors use first author last name et al. (year published)
%% article title  {\it Journal Name} volume #: start page-end page.
%% ie,
% \bibitem{Neuhaus} Neuhaus J-M, Sitcher L, Meins F, Jr, Boller T (1991) 
% A short C-terminal sequence is necessary and sufficient for the
% targeting of chitinases to the plant vacuole. 
% {\it Proc Natl Acad Sci USA} 88:10362-10366.

%% Enter the largest bibliography number in the facing curly brackets
%% following \begin{thebibliography}
\bibliographystyle{unsrt}

%\bibliographystyle{plain}
%\bibliography{}

%%%%%%%%%%%%%%%%%%%%%%%%%%%%%%%%%%%%%%%%%%%%%%%%%%%%%%%%%%%%%%%%

%% Adding Figure and Table References
%% Be sure to add figures and tables after \end{article}
%% and before \end{document}

%% For figures, put the caption below the illustration.
%%
%% \begin{figure}
%% \caption{Almost Sharp Front}\label{afoto}
%% \end{figure}

%% For Tables, put caption above table
%%
%% Table caption should start with a capital letter, continue with lower case
%% and not have a period at the end
%% Using @{\vrule height ?? depth ?? width0pt} in the tabular preamble will
%% keep that much space between every line in the table.

%% \begin{table}
%% \caption{Repeat length of longer allele by age of onset class}
%% \begin{tabular}{@{\vrule height 10.5pt depth4pt  width0pt}lrcccc}
%% table text
%% \end{tabular}
%% \end{table}

%% For two column figures and tables, use the following:

%% \begin{figure*}
%% \caption{Almost Sharp Front}\label{afoto}
%% \end{figure*}

%% \begin{table*}
%% \caption{Repeat length of longer allele by age of onset class}
%% \begin{tabular}{ccc}
%% table text
%% \end{tabular}
%% \end{table*}


 \begin{table*}
\caption{On the full sample, estimated regression. Standard errors are
in brackets. The adjusted $R^2$ is indicated by $\bar{R}^2$ and is
expressed in percentage form.}\label{maintable}
\begin{tabular}{@{\vrule height 10.5pt depth4pt  width0pt}p{1cm}cc|ccc|ccc}
\hline
\\
    & \multicolumn{2}{c}{Regression \ref{simple}}  & \multicolumn{3}{c}{Regression \ref{lagged}} & \multicolumn{3}{c}{Regression \ref{up}} \\
 $q$ & $c$ & $\bar{R}^2 (\%)$ & $c_0$ & $c_1$ &  $\bar{R}^2 (\%)$ & $c_u$ & $c_p$ & $\bar{R}^2 (\%)$ \\
\\
\hline
\\
2 & -0.242 & 1.40 & 9.026 & -11.851 & 3.82 & 12.707 & -6.376 & 4.85 \\
 & (0.805) &  & (1.506) & (1.461) &  & (1.702) & (0.958) &  \\
\\
3 & -0.949 & 0.44 & 10.742 & -14.950 & 3.59 & 13.478 & -7.780 & 3.93 \\
 & (0.933) &  & (1.754) & (1.656) &  & (1.998) & (1.156) &  \\
\\
4 & -0.160 & -0.01 & 12.838 & -16.608 & 3.67 & 14.645 & -7.080 & 3.45 \\
 & (0.978) &  & (1.823) & (1.697) &  & (2.028) & (1.210) &  \\
\\
5 & 0.325 & 0.05 & 14.499 & -18.108 & 3.95 & 15.860 & -6.903 & 3.45 \\
 & (1.051) &  & (1.969) & (1.777) &  & (2.154) & (1.205) &  \\
\\
6 & 0.810 & 0.23 & 16.477 & -20.018 & 4.42 & 17.603 & -6.986 & 3.73 \\
 & (1.108) &  & (2.118) & (1.902) &  & (2.280) & (1.211) &  \\
\\
\hline
\end{tabular}
\end{table*}

\end{document}